\documentclass[twocolumn,PB]{revtex4}
\usepackage{amssymb}

\usepackage{graphicx}
\usepackage{dcolumn}
\usepackage{bm}

\usepackage{graphics}
\begin{document}

\title{Sensitivity dependent model of protein-protein interaction networks}


\author{Jingshan Zhang}
\author{Eugene I. Shakhnovich}
\affiliation{Department of Chemistry and Chemical Biology, Harvard
University, Cambridge, Massachusetts 02138}

\date{\today}

\begin{abstract}

The scale free structure $p(k) \sim k^{-\gamma}$ of protein-protein
interaction networks can be reproduced by a static physical model in
simulation. We inspect the model theoretically, and find the key
reason for the model to generate apparent scale free degree
distributions. This explanation provides a generic mechanism of
``scale free" networks. Moreover, we predict the dependence of
$\gamma$ on experimental protein concentrations or other sensitivity
factors in detecting interactions, and find experimental evidence to
support the prediction.

\end{abstract}
\maketitle
\textbf{1. Introduction} \\ \\
``Scale free" networks have been observed in many areas of
science\cite{Albert} including social science, biology and internet,
where degree distributions follow (albeit noises) the power law form
$p(k) \sim k^{-\gamma}$ within one or two orders of magnitude for
$k$. Here the degree $k$ is the number of links a node has, and
$p(k)$ is the probability of a node to have degree $k$.  An
important scale free network under
experimental\cite{Mewes,Uetz,Ito,Rual} and
theoretical\cite{Deeds,Barabási,Shi,Albert,Jeong,Wagner,Yook,Rzhetsky,Kim}
study is the protein-protein interaction (PPI) network, where a link
between two proteins indicates a large enough binding energy between
them. These studies bare the goal that the topology of PPI networks
could reflect how systems of various proteins have evolved in
biological organisms.

It was pointed out recently that scale free PPI networks could also
result from variation of surface hydrophobicities of proteins.
Starting from an approximately Gaussian distribution of surface
hydrophobicity, the static model successfully produced scale free
networks in computer simulations\cite{Deeds}.

Why can this static model generate scale free networks? As a
counterpart of the simulation results in Ref.\cite{Deeds}, in this
paper we study the model from a theoretical perspective, and reveal
the key reason that the model leads to ``scale free" networks. More
importantly, our numerical and analytical study reveals the
dependence of power $\gamma$ on experimental sensitivity factors,
such as protein concentration, in detection of PPI, and provides a
possible explanation to the observed variation of $\gamma$ in
different high-throughput PPI experiments.
\\

\begin{figure}[ht]
\begin{center}
\includegraphics[width=7.5cm, keepaspectratio]{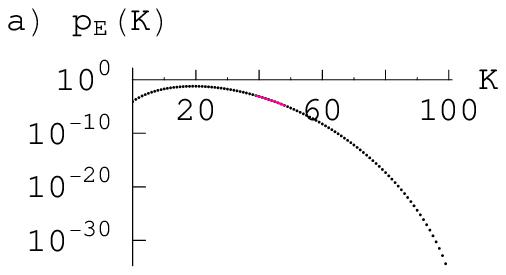} \hfill
\includegraphics[width=7.9cm, keepaspectratio]{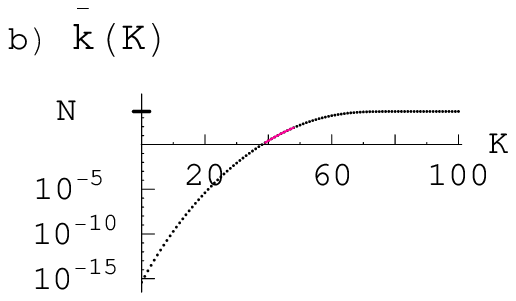}
\end{center}
\caption{(color online). a) Hydrophobicity distribution $p_{_E}(K)$
in Eq.~(\ref{pK1}) for $N\!=\!5000$. The $K$ region in red is the
same as in the inset. b) The dependence of expected degree
$\overline{k}$ upon hydrophobicity $K$, for $K_c\!=\!83$. The range
$1 \le \overline{k} \le 100$ is in red.} \label{figpEK}
\end{figure}

\textbf{2. The static model} \\ \\
Let us first briefly introduce the model proposed by Deeds et
al.\cite{Deeds}. For the compositions of surface residues of yeast
proteins in high-throughput experiments\cite{Uetz,Ito}, the
fractions of hydrophobic residues, noted as $p$, follow a Gaussian
distribution
\begin{equation}
f(p) = {1 \over \sqrt{ 2 \pi \sigma^2 } } e^{-{(p-\overline{p})^2
\over 2 \sigma^2}} \label{fp2}
\end{equation}
with mean value $\overline{p}\!\simeq\!0.2$ and deviation
$\sigma\simeq 0.05$. This results in an approximately Gaussian
distribution of the surface ``stickiness" $K$, and the binding free
energy of two proteins is determined by the sum of their
``stickiness". In a more detailed description, there are $K_i$
hydrophobic residues among the $M$ surface residues on protein $i$,
and $M=100$ is assumed to be a constant for all proteins. The
probability to find a protein with $K$ hydrophobic surface residues
is
\begin{equation}
p_{_E}(K) = \int \!\! dp \,\, f(p) \left(\!\! \begin{array}{ccc}
M \\
K \end{array} \!\!\right)  p^K(1-p)^{M-K}. \label{pK1}
\end{equation}
It can be seen that $p_{_E}(K)$ is close to a Gaussian distribution
(Fig.~\ref{figpEK}a).

The binding of protein $i$ and $j$ is determined by the binding free
energy
\begin{equation}
\Delta G = -(K_i+K_j)F_0+G_{(0)}, \label{binding}
\end{equation}
where $\Delta G$ is negative for a strong binding, $F_0$ is the
change of binding free energy upon burial of each hydrophobic
residue, and $G_{(0)} \simeq 6 kCal/Mol \approx 10 k_BT$ is a
constant value determined by experiments\cite{Janin,Tamura}. In
support of this model, Fig. 3 of Ref.\cite{Janin} showed that
experimental result of binding energies can be described by the sum
of stickiness terms and a constant term. If $K_i+K_j\ge K_c$ the
interaction is experimentally detectable, and the two proteins are
labeled as linked in the PPI network.
\\

\begin{figure}[ht]
\begin{center}
\includegraphics[width=7.5cm, keepaspectratio]{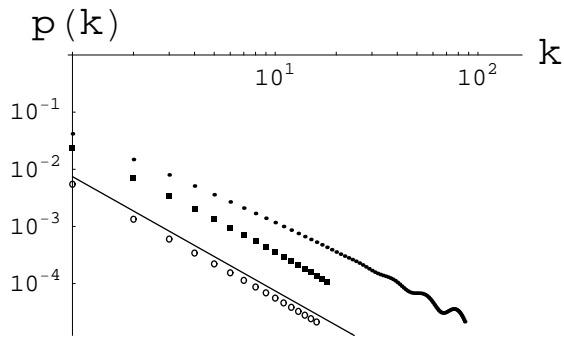}
\end{center}
\caption{ ``Power law" degree distribution $p(k)$ for different
situations, with a solid line indicating slope $\gamma\!=\!2$.
Circles (default): $N\!=\!5000$ and $K_c\!=\!83$; dots: $N\!=\!5000$
and $K_c\!=\!75$; squares: $N\!=\!1000$ and $K_c\!=\!75$. Only data
with $p(k) \ge {1 \over 10N}$ are shown.} \label{figpj}
\end{figure}

\begin{figure}[ht]
\begin{center}
\includegraphics[width=8.5cm, keepaspectratio]{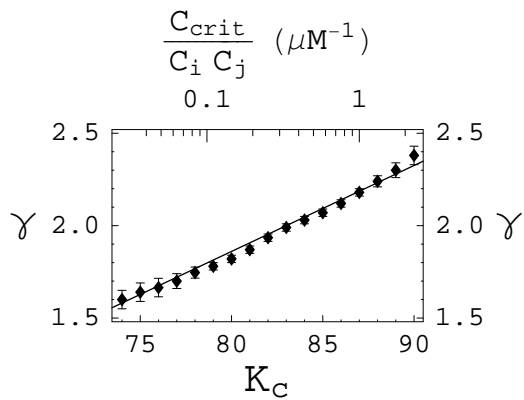}
\end{center}
\caption{Dependence of the power $\gamma$ upon experimental
sensitivity in detecting interactions. $\gamma$ increases with
$K_c$, and $K_c$ is replaced on the top by ${C_{crit}\over C_i C_j}$
from Eq.~(\ref{Kcc}). The error bar at $K_c \le 78$ comes mostly
from undulations. The slight off $p(k=1)$ produces bigger error bar
at $K_c \ge 88$ where there are less $k$ data points. The solid line
is the approximation Eq. (\ref{gammaKc}).} \label{figgammakc}
\end{figure}

\textbf{3. Results and interpretations} \\ \\
\textit{3.1. Degree distributions} \\
We calculate $p(k)$ numerically (see Numerical method for details)
with given values of $N$ and $K_c$, where $N$ is the total number of
proteins in the network, and obtain apparent ``scale free" structure
$p(k)\propto k^{-\gamma}$ (Fig.~\ref{figpj}). We set the default
situation as $N\!=\!5000$ and $K_c\!=\!83$ to fit $\gamma\!=\!2$.
Fig.~\ref{figpj} indicates that the apparent slope $\gamma$
increases with $K_c$, and increases as $N$ decreases. More
explicitly, the dependence of $\gamma$ upon $K_c$ is plotted in
Fig.~\ref{figgammakc}.


Let us interpret these results by analytical approaches. A protein
with hydrophobicity $K$ has a \textsl{pass/fail} line
\begin{equation}
K_{line} =K_c-K\,. \label{passline}
\end{equation}
Proteins with hydrophobicity above $K_{line}$ are linked to it,
while those with hydrophobicity below $K_{line}$ do not. Therefore
the protein with hydrophobicity $K$ has an average degree
\begin{equation}
\!\!\!\overline{k} \simeq N\int_{K_{line}}^{\infty} p_{_E}(K')dK'~~~
\label{jK2}
\end{equation}
In the mean field approximation the degree of the protein $k$ is
just $\overline{k}$, and the degree distribution is
\begin{equation}
\!\!\!p(\overline{k}) = p_{_E}(K) {dK \over d\overline{k}}={
p_{_E}(K) \over N p_{_E}(K_c-K)}\, .   \label{idealpj}
\end{equation}
Beyond mean field approximation its degree fluctuates with deviation
$ \sim \sqrt{\overline{k}}$, which will be addressed later.

Let us restrict the discussion within the mean field approximation
for the moment. We can notice that the experimentally observable
range $1 \!\lesssim\! \overline{k} \!\lesssim\!100$ only covers a
small range of hydrophobicity ($39 \!\lesssim\! K \!\lesssim\! 48$
for the default situation), as indicated by the short red line in
Fig.~\ref{figpEK}b. In this range the hydrophobicity distribution
$p_{_E}(K)$ is very close to exponential, since the short red line
in Fig.~\ref{figpEK}a is nearly straight. So we can use linear
approximation to produce the nearly straight lines in
Fig.~\ref{figpj}. Define
\begin{equation}
a \triangleq -{d\ln p_{_E}(K') \over dK'}|_{K_{line}} \label{adef}
\end{equation}
and
\begin{equation}
b \triangleq -{d\ln p_{_E}(K') \over dK'}|_{K}\,, \label{bdef}
\end{equation}
then Eq.~(\ref{jK2}) give $\overline{k}= e^{a K+const}$, while
Eq.~(\ref{idealpj}) leads to $p(\overline{k})= e^{-(a+b) K+ const}$.
As a result we have $p(\overline{k}) \sim
\overline{k}^{\,-(1+b/a)}$. This is a ``scale free" network with
$\gamma=1+b/a$.

To understand the undulations in $p(k)$ at large $k$ in
Fig.~\ref{figpj}, we must go beyond the mean field approximation and
deal with the fluctuation of degree with magnitude
$\sqrt{\overline{k}}$ for a given $\overline{k}$. Noticing the $K$
values are discrete integers, each $K$ value produces a peak in
$p(k)$, centered at $\overline{k}$ and with width
$\sqrt{\overline{k}}$. Since $\overline{k}$ grows with $K$ almost
exponentially, the distance between nearest neighbor peaks
$\overline{k}(K\!+\!1) \!-\!\overline{k}(K)$ grows linearly with
$\overline{k}$. The undulations emerge at large enough
$\overline{k}$, when the peak distance exceeds the peak width
$\sqrt{\overline{k}}$.

Now we are ready to study the dependence of the slope $\gamma$ on
parameter $K_c$ in Fig.~\ref{figgammakc}. Approximating the
hydrophobicity distribution as Gaussian distribution $\ln
p_{_E}(K)\!\sim\! -(K\!-\! K_0)^2$, where $K_0$ is the most probable
hydrophobicity value, we have
\begin{equation}
\gamma =1+{b \over a} \approx 1+{K_c-K_{line}-K_0 \over
K_{line}-K_0}. \label{gammaKc}
\end{equation}
We find $K_0 \simeq 20$ in Eq.~(\ref{pK1}), and $K_{line} \!\simeq\!
41.5$ is nearly a constant from Eq.~(\ref{jK2}) for typical degree
$k\!\simeq\!5$, then $\gamma$ is a linear function of $K_c$ in
Eq.~(\ref{gammaKc}), and forms a straight line (solid) in
Fig.~\ref{figgammakc}.
\\ \\
\textit{3.2. Dependence on experimental sensitivity} \\
Different $\gamma$ values have been obtained in different PPI
experiments, varying from $\gamma\!\approx \!2.1$ to
$\gamma\!\approx \! 2.5$ \cite{Uetz,Ito,Rual,Jeong,Wagner,Yook}. To
explain this variation, we notice that different experiments might
have different sensitivity in detecting PPI. Indeed, some
interactions detectable in one experiment might be too weak to be
detected in another experiment. An example of factors affecting
experimental sensitivity is protein concentration/level, which is in
turn controlled by gene expression and dependent upon the specific
technique used to detect PPI. Even for the same experiment, the
sensitivity in detecting interactions is actually reduced by setting
a higher standard in identifying PPI, e.g., selecting only highly
repeatable PPI data which effectively correspond to interactions
with high affinity.

Let us study how $\gamma$ depends on these experimental sensitivity
factors. In high-throughput experiments the concentration of
protein-protein complex $C_{ij}$ must be high enough to be detected
\begin{equation}
C_{ij} ={C_iC_j \over C_0} \exp \left(-{\Delta G \over k_BT} \right)
\ge C_{crit}, \label{Kccnew}
\end{equation}
where the binding free energy $\Delta G$ is given by
Eq.~(\ref{binding}), $C_i$ and $C_j$ are the concentrations of
proteins $i$ and $j$ in monomeric form, and the normalization
concentration $C_0=1M$ is the convention. Rewriting this
relationship in the form of association constant, the binding
affinity should be strong enough to be detectable
\begin{eqnarray}
\mathcal{K}_a={1\over C_0} \exp \left[{(K_i+K_j)F_0-G_{(0)}\over
k_BT} \right] \nonumber \\ \,\,\, \ge {1\over C_0} \exp
\left[{K_cF_0-G_{(0)}\over k_BT} \right] ={C_{crit}\over C_i C_j}.
\label{Kcrit}
\end{eqnarray}
Thus the parameter $K_c$ of the model is determined by experimental
protein concentrations
\begin{equation}
K_c = \left[k_BT  \ln \left({C_0C_{crit}\over
C_iC_j}\right)+G_{(0)}\right]/F_0. \label{Kcc}
\end{equation}
To estimate the only unknown parameter $F_0$ in this equation, we
notice that for yeast two hybrid screening technique the PPIs with
binding affinity $\mathcal{K}_a \ge {C_{crit}\over C_i C_j} \simeq 1
\mu M ^{-1}$ are detectable\cite{Estojak}. If we use $\gamma \approx
2.3$ and $K_c=87$ for this threshold binding affinity, we can obtain
an estimate $F_0\approx 0.28 k_BT$. With the help of this value we
can use Eq.~(\ref{Kcc}) to convert the x-axis of
Fig.~\ref{figgammakc} from $K_c$ to experimental variable
${C_{crit}\over C_i C_j}$ (top of Fig.~\ref{figgammakc}).

\begin{figure}[ht]
\begin{center}
\includegraphics[width=7.5cm, keepaspectratio]{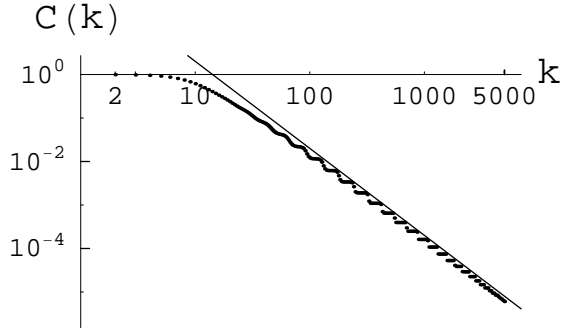}
\end{center}
\caption{The clustering coefficient distribution $C(k)$  for the
default situation
. The solid line indicates slope $-2$.} \label{figCj}
\end{figure}

It can be seen from Fig.~\ref{figgammakc} that lower sensitivity, or
lower $C_iC_j$, leads to higher $\gamma$. This can be realized by
lower protein concentrations through reduced gene expressions, or
selecting only highly repeatable data of detected PPIs. This
prediction is confirmed by Figure 2a of Ref.~\cite{Yook}, which
clearly shows that the core data set of Ito et al.~\cite{Ito},
containing only PPIs identified by at least three independent
sequence tags, generates a steeper degree distribution than the full
Ito data set does. Obviously the Ito core data corresponds to
relatively strong interactions, manifest in high $\mathcal{K}_a$ and
$K_c$. Note that the horizontal dots with $p(k)=1/N$ at high $k$ in
Figure 2a of Ref.~\cite{Yook} should be excluded when fitting the
slope $\gamma$, because they are actually in the $p(k)<1/N$ region
where a few nodes with arbitrary degree $k$ emerge occasionally. On
the other hand, the protein concentrations in the yeast two hybrid
experiments\cite{Uetz,Ito} are not yet available, and the prediction
about dependence of the slope $\gamma$ upon protein concentration
needs verification from future experiments.
\\ \\
\textit{3.3. Clustering coefficient} \\
We also study another important property of networks, clustering
coefficient $C(k)$, and show the numerical result of the model in
Fig.~\ref{figCj}. If a protein is linked to $k$ proteins, the
average number of links between the $k$ proteins, $t(k)$, cannot
exceed $k(k\!-\!1)/2$. Here the averaging includes all possible
realizations. The clustering coefficient is $C(k)\!\triangleq\!
{2t(k) \over k(k-1)} \!\le\! 1$. Similarly to
Ref.\cite{Masuda,Boguna}, we obtain (Fig.~\ref{figCj})
$C(k)\!\simeq\! 1$ at small $k$ and $C(k) \!\sim\! k^{-2}$ at large
$k$. The experimental result\cite{Mewes,Uetz,Ito,Yook} has a similar
shape with slope $\approx2$ for large $k$, and $C(k)$ is smeared
between $1$ and $10^{-1}$ for small $k$. If we attribute the
discrepancy between the model and experiment at small $k$ to false
negatives, the model is in reasonable agreement with experiments.

%
\begin{figure}[ht]
\begin{center}
\includegraphics[width=7.5cm, keepaspectratio]{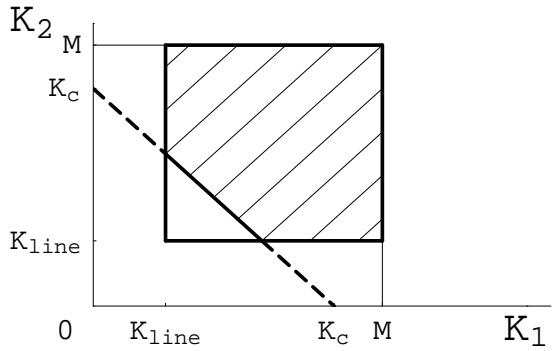}
\end{center}
\caption{The diagram of Eq.~(\ref{cK0}) to find the behavior of
clustering coefficient $C(k)$. The numerator is the integral over
the shadowed region, while the denominator is the integral over the
square region.} \label{figarea}
\end{figure}
A physical picture is helpful to interpret this result. As mentioned
above, if there are the $k$ proteins linking to the same protein,
their hydrophobicity exceed $K_{line}$, while the hydrophobicity of
all other proteins are below $K_{line}$. The mean field relationship
between $K_{line}$ and $k$ is Eq.~(\ref{jK2}). If we have $K_{line}
\!\ge\! K_C/2$ at a small degree $k$, then the most hydrophobic $k$
proteins are all connected, and $C(k)$ is $1$. At large enough $k$,
however, $K_{line}\!<\!K_c/2$ and not all proteins above $K_{line}$
are linked to each other. Then the clustering coefficient is
determined by
\begin{equation}
C(k)\!
\simeq\!\frac{\!\int_{\!K_{lin\!e}}^M\!\!\!dK_{\!1}\!\!\int_{\!M\!ax
\{\!K_{\!c}-\!K_{\!1},K_{lin\!e}\}}^M\!\!dK_{\!2}
p_{_E}(K_{\!1}\!)p_{_E}(K_{\!2}\!)}{\int_{K_{line}}^MdK_1\int_{K_{line}}^MdK_2
\,p_{_E}(K_1)\,p_{_E}(K_2)}. \label{cK0}
\end{equation}
The denominator is proportional to $\overline{k}^{\,-2}$ according
to Eq.~(\ref{jK2}). It corresponds to the square region between
$K_{line}$ and $M$ in Fig.~\ref{figarea}. The numerator,
corresponding to the shadowed region in Fig.~\ref{figarea}, is
dominated by the region near the cutting line $K_1+K_2=K_c$, because
$p_{_E}(K)$ is nearly a sharp exponential function. Hence the
numerator scales as the length of the cutting line, $K_c-2K_{line}
\propto \ln \overline{k}+const$. Therefore, in agreement with Boguna
\textit{et al.}~\cite{Boguna}, the numerator is a slow function of
$k$ compared to the denominator, and the clustering coefficient
scales as $C(\overline{k}) \sim \overline{k}^{\,-2}$ at large
$\overline{k}$. And at small $\overline{k}$ the square is totally in
the shadow, leading to $C(\overline{k}) \sim 1$. The step like shape
of $C(k)$, however, comes from the discreteness of integer $K$
values. \\


%
%

\textbf{4. Numerical methods} \\ \\
We calculate $p(k)$ as an average of all possible realizations. The
calculation is done with integer $K'$ and without mean field
approximation. We ignore the unimportant difference between $N$ and
$N+1$ for large enough $N$. The exact form of Eq. (\ref{jK2}) is
\begin{equation}
\overline{k} = \!\! N\cdot \sum _{K'=Max\{K_c-K,0\}} ^M
\!\!p_{_E}(K')~, \label{jK}
\end{equation}
and the degree distribution is
\begin{equation}
p(k) =\sum_{K=0}^M p_{_E}(K) \left(\!\! \begin{array}{ccc} N \\ k
\end{array} \!\!\right) ({\overline{k} / N} )^{k}
(1-{\overline{k} / N} )^{N-k}.  \label{pj}
\end{equation}

Instead of mean field result Eq.~(\ref{cK0}), the clustering
coefficient is calculated as
\begin{eqnarray}
C(k) =\sum_{K=0}^{M-1} \{w(K) /[\sum_{K_1,K_2=K}^{M}
p_{_E}(K_1)p_{_E}(K_2)]~~~~~~ \nonumber \\
\times[\!\!\!\!\!\!\sum_{K_1,K_2=K}^{M} \!\!\!\!\!\!\!\!
p_{_E}(K_1)p_{_E}(K_2) \theta(K_1+K_2-K_c+1/2)]\}
  \label{cj1}
\end{eqnarray}
where
\begin{equation}
\theta(K) =   \left\{\!\! \begin{array}{ccc}
1 &  \,\,\,\,\,\,K > 0 \\
0 &  \,\,\,\,\,\,K<0 \end{array} \!\!\right.   \label{theta}
\end{equation}
is the usual Heaviside step function, and
\begin{equation}
\!w(K)\!=\!\! \left(\!\!\! \begin{array}{ccc} N \\ k
\end{array} \!\!\!\right)\![\!\!\sum_{K'=\!K}^{M}\!\!\!p_{_E}(K')]^{k}
\{[\!\!\sum_{K'=0}^{K\!-\!1}\!\!p_{_E}(K')]^{\!N\!-\!k}\!-\![\!\!\sum_{K'=0}^{K\!-\!2}\!\!p_{_E}(K')]^{\!N\!-\!k}\}
\label{wK}
\end{equation}
is the probability that $k$ proteins have hydrophobicity $\ge K$
while the maximum hydrophobicity of the rest $N-k$ proteins is
$K-1$.
\\

\textbf{5. Conclusion and outlook} \\ \\
We study a static physical model to explain scale free PPI networks.
We notice that the experimentally observable part of degree
distribution covers a limited range (from $k=1$ to $k<100$), and
corresponds to a small range of hydrophobicity. The hydrophobicity
distribution $p_{_E}(K)$ in this small range is close enough to an
exponential distribution. Therefore a linear approximation leads to
the ``scale free" degree distribution $p(k)\!\sim\! k^{-\gamma}$,
with $\gamma$ dependent on the threshold parameter $K_c$ and network
size $N$. In experiments $K_c$ depends on the sensitivity factors,
such as protein concentration, in detection of PPI. Our result
provides a possible interpretation to the difference in experimental
$\gamma$ values, and predicts the dependence of $\gamma$ on
experimental sensitivity factors. This prediction is supported by
the slope change\cite{Yook} when comparing Ito data set and Ito core
data set\cite{Ito}, and dependence of $\gamma$ on protein
concentrations needs experimental verification in future. The
distribution of another network property, clustering coefficient,
produced in the model is also in reasonable agreement with that of
experiment\cite{Yook} and previous theoretical
descriptions\cite{Masuda,Boguna}.

The hydrophobicity distribution in the physical model has been
arranged to reflect the reality in a simplified way. While the real
distribution of protein ``stickiness" can be somewhat different from
it, the generation of ``scale free" network will not be sensitive to
the difference. More generally, ``scale free" degree distributions
can be also produced by many smooth distributions of hydrophobicity,
such as binomial, Gaussian, Poisson distributions and their
modifications. This can be one of the reasons that scale free (in a
limited range) networks are so widely observed.

A major part of PPI networks is obtained by the high-throughput
yeast two hybrid screening\cite{Uetz,Ito}. This technique often
produce a large fraction of false positives\cite{Deane} which do not
correspond to any real biological function, while real functional
interactions presumably constitute a smaller portion of the detected
result. While functional PPIs may involve formation of additional
hydrogen bonds and salt bridges to obtain adequate binding affinity,
these nonfunctional PPIs have not been evolutionarily selected and
are formed primarily due to hydrophobic effect. In this model we
show that a simple static network of nodes with different
``stickiness" can readily appear to be scale free. To this end, we
use Eq.~(\ref{binding}) because the nonfunctional PPIs are just
random interfaces between two proteins without experiencing the
evolutionary design of pairwise interface patterns. Moreover, this
model could be used to extract information of nonfunctional
interactions between unrelated proteins which randomly encounter in
a real cell, and such information is in turn important in probing
the general principles for cells to organize proteins in a cell.
Namely, the stronger nonfunctional interactions, the more unrelated
proteins interfere with each other, and the less protein types can
coexist. Hence the nonfunctional interactions can limit the proteome
size of a single cellular organism. Interesting related questions
include the change of how much living cells have to do in
constricting the nonfunctional interactions in the course of protein
evolution, as well as the impact of higher temperature for
thermophile organisms.

If the distribution of ``stickiness" is simply an exponential
function,  $\ln p_{_E}(K) \sim -K$, the model is simplified to $a=b$
and $\gamma=2$. This reduced simple situation would then be in
complete agreement with one of the mathematical examples of networks
briefly mentioned in by Caldarelli et al.\cite{Caldarelli}, which
has been applied to realistic networks such as gene regulation
network\cite{Balcan}. Our finding indicates that this simple
mathematical form\cite{Caldarelli} have more important impacts to
systems in reality. Indeed, it is reasonable to expect distribution
of many qualities, such as annual personal income and eagerness to
learn knowledge, to be fitted by an approximately exponential
distribution at least in some short range, and with suitable
arrangements power law distribution might emerge.

Masuda et al.\cite{Masuda} followed the suggestion of Caldarelli et
al.\cite{Caldarelli} and studied essentially similar models to ours.
But they did not relate the mathematical models to real systems.
More importantly, they emphasize that the slope $\gamma\!=\!2$ is
universal, while the slope in our study not only deviates from $2$,
but also dependents on experimental properties such as expression
levels of proteins.

In contrast to this static model, most models of PPI networks focus
on the development history of the network through gene
duplications\cite{Rzhetsky,Kim}, which is similar to ``preferential
attachment" in growing networks\cite{Barabási}. It was
found\cite{Kim} that the network structure of the gene duplication
model analytically approaches scale free\cite{Kim} at $k
\!\rightarrow \!\infty$ if links of new nodes should be deleted by a
probability larger than $1/2$, and the degree distribution is
comparable with experiments. Our approach serves as an alternative
way to obtain ``scale free" PPI network. Further experiments, such
as systematic study of dependence of apparent power $\gamma$ on gene
expression level, or other measures of protein concentration, will
help clarify whether the static model or gene duplication mechanism
is mainly responsible for the observed scale free nature of PPI
networks.
\\

\textbf{Glossary} \\ \\
\textit{Protein-protein interaction network.} A network of many
types of proteins of an organism; each type of protein is a node in
the network. Two nodes are labeled as linked if the two types of
proteins can interact with each other with sufficient
affinity.\\ \\
\textit{Degree.} The number of links a node has in the network. If a
node in the protein-protein interaction network has degree $k$, this
protein can interact with $k$ other types of proteins. \\ \\
\textit{Scale free network.} In such a network, the number of nodes
with degree $k$ decreases with $k$, and the dependence is a power law
function.\\ \\
\textit{Yeast two hybrid.} A molecular biology technique used to
discover protein-protein interactions by testing for physical
interaction/binding between two proteins, respectively. This
technique is able to test interactions between a large amount of
proteins
rapidly (so called high-throughput screening).\\ \\
\textit{sensitivity in detecting interactions.} Only strong enough
interactions between proteins are identified as ``interacting"
pairs. If the sensitivity in detection becomes higher, slightly
weaker interactions becomes detectable, and more interactions are
detected.  \\ \\
\textit{Surface hydrophobicity.} The fraction of hydrophobic amino
acids among the amino acids on the surface of a protein. If
hydrophobic amino acids are buried either in formation of a protein
or in formation of a protein-protein complex, they are not in
contact with water any more, and thus lowers the total free energy.
Hydrophobic effect is important in the interaction of proteins,
especially in
non-functional interactions. \\

\textbf{Acknowledgement}: This work is supported by NIH.



\end{document}